\documentclass[aps,pra,twocolumn,showpacs]{revtex4}
\usepackage[]{graphicx}
\usepackage{bm}
\usepackage{amsmath}
\usepackage{amssymb}

\providecommand{\bra}[1]{\langle #1 \rvert}
\providecommand{\ket}[1]{\lvert #1 \rangle}

\begin{document}

\title{Three-qubit topological phase on entangled photon pairs}

\author{Markus Johansson$^{1}$, Antonio Z. Khoury$^{2}$, Kuldip Singh$^{1}$
and Erik Sj\"oqvist$^{1,3}$}

\affiliation{$^{1}$
Centre for Quantum Technologies, National University of Singapore, 3 Science Drive 2, 117543 Singapore, Singapore}
\affiliation{$^{2}$
Instituto de F\'\i sica, Universidade Federal Fluminense,
24210-346 Niter\'oi - RJ, Brazil}
\affiliation{$^{3}$
Department of Quantum Chemistry, Uppsala University, Box 518, SE-751 20 Uppsala, Sweden}

 
\begin{abstract}
We propose an experiment to observe the topological phases associated with cyclic evolutions, 
generated by local SU(2) operations, on three-qubit entangled states prepared on different 
degrees of freedom of entangled photon pairs. The topological phases reveal the nontrivial 
topological structure of the local SU(2) orbits. We describe how to prepare states showing different topological phases, and discuss their relation to entanglement. In particular, the presence of a $\pi/2$ phase shift is a signature of genuine tripartite entanglement in the sense that it does not exist for two-qubit systems.

\end{abstract}
\vskip2pc 

\maketitle

\section{Introduction}

Topological phases of quantum systems that evolve in topologically nontrivial spaces,  
have attracted considerable attention in a wide variety of subdisciplines in modern physics. 
Perhaps the most well-known example of such a topological quantity is the Aharonov-Bohm 
phase acquired by a charged particle that encircles a shielded magnetic flux line 
\cite{aharonov59}. This phase depends only on the winding number of the particle's 
path around the impenetrable region of magnetic flux, but it is insensitive to 
perturbations of the path.

A topological phase acquired by a pair of entangled qubits undergoing cyclic local unitary evolution 
has been discovered \cite{remy2}. The topological interpretation of this phase relies 
on the relation between two-qubit states and the rotation group SO(3) \cite{milman}. This is 
perhaps most clearly seen in the case of the maximally entangled states. These states are in 
one-to-one correspondence with the points of real projective space $S^3/Z_2 \sim {\textrm{SO(3)}}$ 
\cite{remy,kus01}. The two possible topological phases $0$ and $\pi$ can be associated with the 
two homotopy classes of loops in SO(3). In other words the accumulated phase 
is not affected by continuous deformations of path of the cyclic evolution. The topological 
two-qubit phase has been observed in spin-orbit transformations on a laser beam \cite{topoluff} 
and in a nuclear magnetic resonance setting \cite{nmr}. 

The notion of topological phase has been extended to pairs of entangled higher-dimensional 
quantum systems \cite{fracuff}. These phases are integer multiples of $2\pi /d$, where $d$ is 
the Hilbert space dimension of each subsystem. Thus, for such objects, fractional values may 
occur. The topological phase for a given cyclic local SU($d$) evolution of a state $\ket{\psi} = 
\sum_{k,l=1}^d \alpha_{kl} \ket{kl}$ is restricted by the invariance of the determinant 
$\det \alpha_{kl}$ of the coefficient matrix. Since $\det \alpha_{kl} = 0$ for product states, 
the topological phase is only well-defined in the presence of entanglement. 

Recently, the notion of topological phase has been extended to $N$-qubit systems 
\cite{multiqubit}. These multi-qubit phases may take fractional values for $N \geq 3$. 
The number of possible values increases rapidly with the number of qubits. All possible values for up to $N=7$ 
have been found using a combinatorial algorithm \cite{multiqubit}. Furthermore, a relation 
between the topological phases and the degree of nonzero polynomial entanglement 
invariants has been conjectured \cite{multiqubit}. As an example of such a relation, the possible topological 
phases $0,\pi/2,\pi$, and $3\pi/2$ for $N=3$ can be linked to 
multipartite entanglement in the sense that three-tangle is a polynomial invariant of 
degree $n=4$, namely the hyperdeterminant in the coefficient matrix $\alpha_{klm}$ 
\cite{coffman00}. This implies that the allowed topological phases are indeed restricted 
to integer multiples of $2\pi /n=\pi/2$. 

In order to realize a multiple qubit system in a photonic device, one may combine 
different degrees of freedom that can be manipulated independently. Numerous 
experiments have employed polarization and orbital angular momentum (OAM) 
to implement controlled operations \cite{qplates1,qplates2,qplates3,cnot,bb84} and 
spin-orbit Bell inequality \cite{bell1,bell2}. 
Here, we propose an experiment to measure the topological phases for $N=3$, in 
qubits encoded on photon pairs produced by spontaneous parametric down conversion (SPDC). 
Each photon carries a polarization and orbital degree of freedom. 
The three qubits are encoded in the orbital part of the signal photon and the two polarizations, 
by projecting the orbital part of the idler photon on a well-defined Laguerre-Gaussian mode. 
In this way, we demonstrate different three-qubit states that acquire the different three-qubit 
phases by employing local SU(2) transformations in Franson loop interferometers on each 
photon. The observed phases would be a signature of the local orbits and thereby a non-trivial 
signature of multipartite entanglement. 

The outline of the paper is as follows. The theory of topological three-qubit phases arising 
in local SU(2) evolution is described in Sec. II. Sections III-V contain the experimental 
setup, where the generation of three different types of three-qubit states are described in 
Sec. III, the measurement of topological phases is described in Sec. IV, and examples of evolutions 
that reveal the topological phases is given in Sec. V. The paper ends with the conclusions.

\section{Three-qubit topological phase structure}
When considering interconvertibility of three-qubit states under stochastic local operations 
and classical communication (SLOCC), the genuinely tripartite entangled states fall into two 
classes \cite{dur}. These classes are termed the GHZ-class and the W-class after their 
representatives, the GHZ state and the W state. 

By considering interconvertibility 
under local unitary transformations the two SLOCC-classes can be further divided into local 
unitary classes, or in other words, orbits of the group of local unitary transformations.
The structure of such an orbit constitutes a qualitative description of the entanglement of the 
states belonging to the orbit. This is the most detailed description of the entanglement properties 
that can be given \cite{sud}. Since the action of the $U(1)$ group is a trivial global phase shift, 
it is sufficient to consider the local SU(2)-orbits to study entanglement properties. 

The structure of the SU(2)-orbits of entangled three-qubit states has been studied by Carteret and 
Sudbery in Ref. \cite{car}. In particular it was shown that the local SU(2)-orbit of a GHZ state 
$|\psi_{ghz}\rangle=\frac{1}{\sqrt{2}}(|+++\rangle+|---\rangle)$, where $|+\rangle$ and 
$|-\rangle$ are orthogonal states, is quadruply connected. The four different homotopy classes 
of cyclic evolutions correspond to the four different accumulated phases, 
$0,\frac{\pi}{2},\pi$ and $\frac{3\pi}{2}$. Since these are the only phases allowed for a state with 
nonzero three-tangle it follows that the quadruple connectedness is related to the tripartite 
entanglement measured by the three-tangle. A $\frac{\pi}{2}$ phase shift cannot be generated in a 
two-qubit system and is therefore a measurable quantity that indicates the presence of tri-partite 
entanglement.

Four different topological phases is in fact not the most common topological phase structure for local 
SU(2)-orbits belonging to the GHZ SLOCC-class. Using the canonical form of three-qubit 
states of Carteret {\it et al.} \cite{higu}, it can be seen that the set of local SU(2) orbits that exhibit four topological phases forms a subset of the local SU(2) orbits of the GHZ SLOCC-class 
parameterized by four real parameters, while the full set of local SU(2) orbits is parameterized 
by four real and one complex parameter. The states of this subset can, up to local SU(2) operations, be written 
on the form
\begin{eqnarray}
|X_{a,b,c,d}\rangle &=& a|+++\rangle+b|+--\rangle 
\nonumber\\
 & & + c|-+-\rangle + d|--+\rangle, 
\end{eqnarray}
where 
$a,b,c,d\in\mathbb{C}\backslash\{0\}$ such that $|a|^2+|b|^2+|c|^2+|d|^2=1$.
We will refer to this class of states as the X-class. 
A distinguished member of this class 
is the three-qubit state $|\psi_{X}\rangle$ for which $a=b=c=d=\frac{1}{2}$ termed the 
three-qubit X state in Ref. \cite{ost}. This state is maximally entangled in the sense that all 
reduced density operators for the individual qubits are proportional to the identity. Note that the 
X state can be brought to the GHZ state by application of a Hadamard transformation on 
each qubit. Hence, these two states are entangled in exactly the same way.  

The states in the GHZ SLOCC-class that do not fall in the X-class have only two different topological phases. 
Since the X-class is a lower dimensional subset, a generic 
state in the GHZ SLOCC-class is of this kind.
An example of such a state, with a doubly connected 
local SU(2) orbit, is a biased GHZ state $|\psi_{bghz}\rangle = 
\alpha|+++\rangle+\beta|---\rangle$, where $|\alpha|\neq|\beta|$ and $|\alpha|^2+
|\beta|^2=1$ \cite{car}. The two homotopy classes of cyclic evolutions correspond to 
the accumulated phases $0$ and $\pi$.

The X state and a biased GHZ state thus represents the two different topological phase structures 
present in the GHZ SLOCC-class. The remaining three-qubit states with genuine tripartite 
entanglement belong to the W SLOCC-class, and have either the topological phases $0$ 
and $\pi$, or no topological phases at all. We would thus not see any other sets of topological 
phases by studying states in the W class.

This paper is concerned with three-qubit systems encoded in the polarization and orbital angular 
momentum (OAM) states of 
photons. We will be describing the polarization states in a basis of right and left circular 
polarization states $|+\rangle$ and $|-\rangle$ or alternatively in a basis of horizontal 
and vertical polarization states $|H\rangle$ and $|V\rangle$. The relation between these 
basis vectors is given by $|\pm\rangle=\frac{1}{\sqrt{2}}(|H\rangle\pm{i}|V\rangle)$. 
The OAM states will be described in 
terms of a basis of Laguerre-Gaussian modes of first order ${\textrm{LG}}_{1,0}$ and 
${\textrm{LG}}_{-1,0}$, denoted $|+\rangle$ and $|-\rangle$ similarly to the circular 
polarization states, or in a basis of the Hermite-Gaussian first order modes 
${\textrm{HG}}_{1,0}$ and ${\textrm{HG}}_{0,1}$, denoted $|h\rangle$ and $|v\rangle$ 
similarly to the linear polarization states. The relation between these bases is given by 
$|\pm\rangle=\frac{1}{\sqrt{2}}(|h\rangle\pm{i}|v\rangle)$.

It is useful to note that the X state prepared in a basis of circular polarization states and 
Laguerre-Gaussian modes is the GHZ state, up to a relative phase factor $-i$ of the two 
terms, in a basis of horizontal and vertical polarization states and Hermite-Gaussian modes.
For example, if the X state in the $\{|+\rangle,|-\rangle\}$ basis has been encoded in the 
polarization and OAM states of a photon pair, such that the first and last qubit are encoded 
in polarization states and the middle in the OAM state of one of the photons, the same state 
in the $\{|H\rangle,|V\rangle\}$ and $\{|h\rangle,|v\rangle\}$ basis would be 
$\frac{1}{\sqrt{2}}(|HhH\rangle-i|VvV\rangle)$.

We will consider the X state, the GHZ state, and a biased GHZ state in the $\{|+\rangle,|-\rangle\}$ 
basis since this allows us to implement cyclic local SU(2) evolutions that reveal the topological 
phases and lie completely within the set of operators that diagonalize in the $\{|+\rangle,|-\rangle\}$ 
basis. Considering the X state there are evolutions in each homotopy class that diagonalize in the 
$\{|+\rangle,|-\rangle\}$ basis, and thus allows all possible topological phases to be observed. 
This is true also for the biased GHZ state.

\section{Quantum State Preparation}

Our experimental proposal is based on the spontaneous parametric down conversion 
(SPDC) source 
of entangled photons first demonstrated in Ref. \cite{kwiat}, and later used in 
other experiments \cite{posinterf,imagepol}. There, two adjacent nonlinear 
crystals cut for type I phase match are spatially oriented with their 
optical axis mutually orthogonal. Starting from a linearly polarized laser, a 
quarter waveplate (QWP-p) can be used to produce a circularly polarized pump, and 
generate pairs of polarization entangled photons of the kind
\begin{equation}
\ket{\psi_{pol}}=\frac{\ket{HH}-i\ket{VV}}{\sqrt{2}}\;,
\label{psipol}
\end{equation}
where the first term on the right hand side comes from the $V$ component of the pump 
while the second one comes from the $H$ component. 

In order to 
realize the three-qubit system, we may add the orbital angular momentum (OAM) 
quantum state of the photon pair \cite{miles}. As already demonstrated 
\cite{mair,uffufrj,uffusp}, the spatial correlations imposed by the 
phase match condition in parametric down conversion are manifested in the OAM 
transfer from the pump to the down converted photons, giving rise to an OAM 
entangled state of the form
\begin{equation}
\ket{\psi_{oam}}=\sum_{m} C_m \ket{m\, ,\,l-m}\;,
\label{psioam}
\end{equation}
where $m$ and $l$ are the topological charges of signal and pump photons, 
respectively. Then, OAM conservation imposes that the added topological 
charge of signal and idler equals that of the pump, leading to a superposition 
of all components compatible with this condition. The probability amplitudes 
$C_m$ associated with a particular OAM partition is proportional to the 
spatial overlap between signal, idler, and pump transverse modes \cite{torres}. 
Now, the three-qubit realization can be achieved by pumping the SPDC source 
with a Laguerre-Gaussian mode with $l=+1$ and detecting the idler photon 
with a single mode fiber (SMF) that admits only the $l-m=0$ component. 
Then, coincidence measurements should be obtained only for signal photons 
with $m=+1$. Therefore, the postselected spin-orbit quantum state is 
\begin{equation}
\ket{\psi_{SO}}=\ket{\psi_{pol}}\otimes\ket{+,0}\;.
\label{psiSO}
\end{equation}

Since the subspace of first order paraxial modes have a 
qubit structure \cite{poincare}, we can now encode two qubits on the 
signal photons, namely their polarization and OAM, and a single 
qubit on the idler polarization. From now on, we shall omit the 
idler OAM since no operations other than detection filtering will 
be performed in this degree of freedom. Therefore, the initial 
three-qubit state generated is
\begin{equation}
\ket{\psi_{1}}=\frac{\ket{H+H}-i\ket{V+V}}{\sqrt{2}}\;,
\label{psi1}
\end{equation}
where we have grouped together the signal degrees of freedom. 
Now we shall discuss separately the two entangled three-qubit states of 
interest. We further show how to prepare certain product
states that are used to investigate the role of entanglement in the
topological phase measurements.

\subsection{X State}

\begin{figure}
\begin{center} 
\includegraphics[scale=.6]{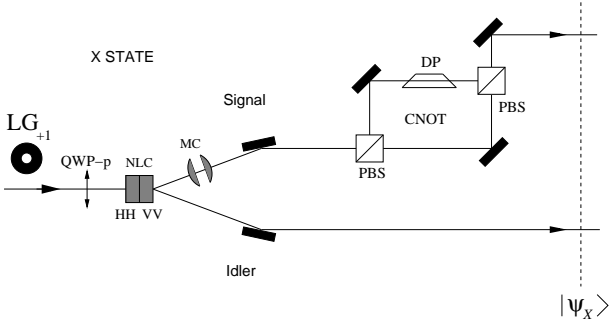}\\
\end{center} 
\caption{X state preparation setup.}
\label{xsetup}
\end{figure}

First, we will see how to produce the three-qubit quantum state 
showing the $\pi/2$ topological phase. The proposed setup is 
sketched in Fig. \ref{xsetup}. In order to simply understand 
the setup, it is useful to recall that the X state in the 
$\{\ket{+},\ket{-}\}$ basis corresponds to a GHZ state in 
the $\{\ket{H},\ket{V}\}$ basis. Therefore, following the 
setup, we shall be seeking for this state. First, an 
astigmatic mode converter can be used to transform the 
signal LG mode to a horizontal first order HG mode 
\cite{converter}, giving
\begin{equation}
\ket{\psi_{1}}\rightarrow\ket{\psi_{2}}=
\frac{\ket{HhH}-i\ket{VhV}}{\sqrt{2}}\;.
\label{psi2}
\end{equation}
This state could also be produced by pumping the crystals with 
the first order Hermite-Gaussian mode $h$, still filtering the idler 
with the single mode fiber. In this case, the signal mode with optimal 
spatial overlap with pump and idler is also $h$. This would exempt the 
use of the mode converter, making the system alignment considerably 
easier. 

Then, a spin-orbit controlled NOT (CNOT) gate is used to flip 
the signal HG mode conditioned to its polarization. The CNOT gate 
is a Mach-Zehnder interferometer with input and 
output polarizing beam splitters (PBS). A Dove prism (DP) 
oriented at $45^o$ and inserted in the $(V)$ arm makes the transverse 
mode conversion $\ket{h}\rightarrow\ket{v}$ on this arm. 
After the CNOT gate the three-qubit 
quantum state becomes the desired X state:
\begin{eqnarray}
\ket{\psi_{X}}&=&\frac{\ket{HhH}-i\ket{VvV}}{\sqrt{2}}
\nonumber\\
&=&\frac{\ket{+++}+\ket{+--}+\ket{-+-}+\ket{--+}}{2}\;.
\label{psi3}
\end{eqnarray}
%

\subsection{Biased GHZ state}

\begin{figure}
\begin{center} 
\includegraphics[scale=.6]{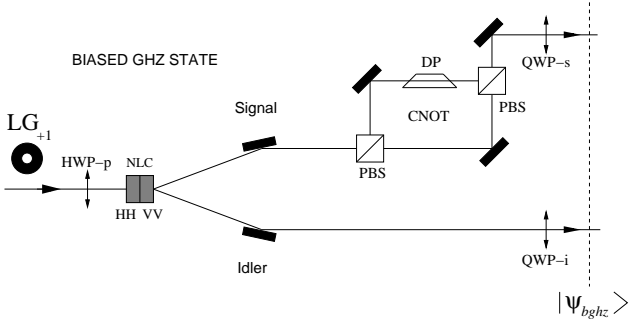}\\
\end{center} 
\caption{Biased GHZ preparation setup.}
\label{ughzsetup}
\end{figure}

In order to produce the biased GHZ state showing only topological 
phase $\pi$, the setup shown in Fig. \ref{ughzsetup} can be used. 
First, a half waveplate (HWP-p) with a suitable orientation is placed 
on the pump laser to set its polarization to produce the partially 
entangled state 
\begin{equation}
\ket{\psi^{\prime}_{pol}}=\alpha\ket{HH}+\beta\ket{VV}\;,
\label{psipolprime}
\end{equation}
so that the initial three qubit state will be
\begin{equation}
\ket{\psi^{\prime}_{1}}=\alpha\ket{H+H}+\beta\ket{V+V}\;.
\label{psi1prime}
\end{equation}
With the astigmatic mode converter removed, the transformation 
$\ket{+}\rightarrow\ket{-}$ is performed in the $(V)$ arm of the 
CNOT gate, giving
\begin{equation}
\ket{\psi^{\prime}_{2}}=\alpha\ket{H+H}+\beta\ket{V-V}\;.
\label{psi2prime}
\end{equation}
Now, two quarter wave plates inserted on signal (QWP-s) 
and idler (QWP-i) paths, make the polarization transformations 
$\ket{H}\rightarrow\ket{+}$ and $\ket{V}\rightarrow\ket{-}$ 
needed to produce the desired biased GHZ state
\begin{equation}
\ket{\psi_{bghz}}=\alpha\ket{+++}+\beta\ket{---}\;.
\label{psi3prime}
\end{equation}
%

\subsection{Product states}

In order to investigate the role of entanglement in the topological phase 
measurements, it is important to compare the quantum states discussed above 
with product states that are equivalent to the X state and the biased 
GHZ state in what regards the single qubit probabilities. 
For example, the product state 
\begin{eqnarray}
\ket{\psi_{prod}}&=&
\frac{\ket{H}+\ket{V}}{\sqrt{2}}\otimes
\frac{\ket{h}+\ket{v}}{\sqrt{2}}\otimes
\frac{\ket{H}+\ket{V}}{\sqrt{2}}\nonumber\\
&=&\frac{e^{-i\frac{\pi}{4}}\ket{+}+e^{i\frac{\pi}{4}}\ket{-}}{\sqrt{2}}\otimes
\frac{e^{-i\frac{\pi}{4}}\ket{+}+e^{i\frac{\pi}{4}}\ket{-}}{\sqrt{2}}\nonumber\\
&&\otimes
\frac{e^{-i\frac{\pi}{4}}\ket{+}+e^{i\frac{\pi}{4}}\ket{-}}{\sqrt{2}}
\label{psiprod}
\end{eqnarray}
has the same probability distribution as the X state for each individual 
degree of freedom in both the $\{|H\rangle,|V\rangle\}$ basis and the 
$\{|+\rangle,|-\rangle\}$ basis.

\begin{figure}
\begin{center} 
\includegraphics[scale=.6]{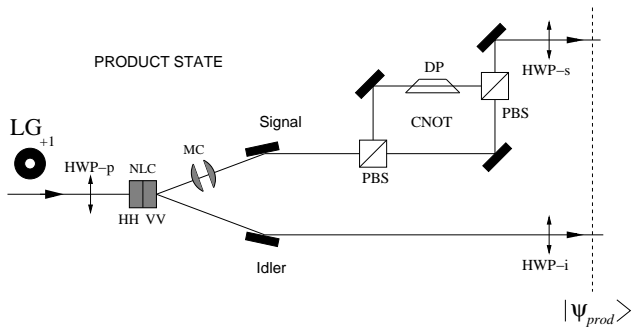}\\
\end{center} 
\caption{Product state preparation setup.}
\label{prodsetup}
\end{figure}

This state is readily prepared by the setup shown in Fig. \ref{prodsetup}
when the pump polarization is set to $V$ and the down converted photons 
are created at the product state $\ket{H+H}$. In the signal arm, the mode 
converter is then oriented to make the transformation $\ket{+}\rightarrow 
(\ket{h}+\ket{v})/\sqrt{2}$, and the $H$ polarization passes unaffected 
through the CNOT gate. Then, two half-waveplates can be used to set signal 
(HWP-s) and idler (HWP-i) polarizations to $(\ket{H}+\ket{V})/\sqrt{2}$, 
thus producing $\ket{\psi_{prod}}\,$. 

A product state with the same probability distributions as the biased 
GHZ state in the $\{|+\rangle,|-\rangle\}$ basis could be
\begin{eqnarray}
\ket{\psi^{\prime}_{prod}}&=&
\left(\widetilde{\alpha}\ket{H}+\widetilde{\beta}\ket{V}\right)\otimes
\left(\widetilde{\alpha}\ket{h}+\widetilde{\beta}\ket{v}\right)
\nonumber\\
&\otimes&
\left(\widetilde{\alpha}\ket{H}+\widetilde{\beta}\ket{V}\right)\nonumber\\
&=&\frac{\alpha\ket{+}+\beta\ket{-}}{\sqrt{2}}\otimes
\frac{\alpha\ket{+}+\beta\ket{-}}{\sqrt{2}}
\nonumber\\
&\otimes&
\frac{\alpha\ket{+}+\beta\ket{-}}{\sqrt{2}},
\label{psiprodprime}
\end{eqnarray}
where $\widetilde{\alpha}=\frac{\alpha+\beta}{\sqrt{2}}$ and 
$\widetilde{\beta}=i\frac{\alpha-\beta}{\sqrt{2}}$. $\ket{\psi^{\prime}_{prod}}$ 
can be produced in the same way as $\ket{\psi_{prod}}$, but with 
suitable settings of the mode converter and the HWPs in order to provide 
the coefficients $\widetilde{\alpha}$ and $\widetilde{\beta}$. 
Both $\ket{\psi_{prod}}$ and $\ket{\psi^{\prime}_{prod}}$ could also 
be produced by tailoring the pump mode in order to optimize the 
spatial overlap between pump, idler, and the desired signal mode, 
without the use of a mode converter on the signal arm. 

The role played by entanglement 
in the topological phase evolution can be investigated with two-photon 
interferometry, as we shall see in section \ref{numerical}. The interference 
patterns produced by entangled states are clearly distinguished from those 
expected for product states. 

\section{Topological phase measurement}

\begin{figure}
\begin{center} 
\includegraphics[scale=.6]{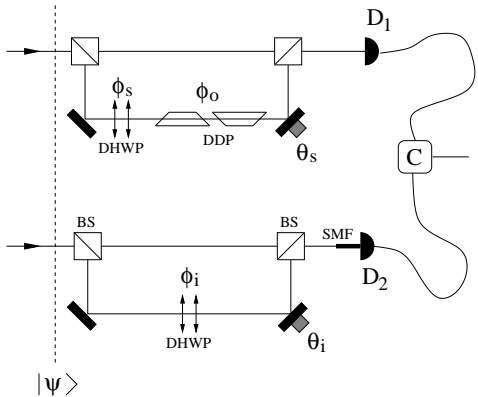}\\
\end{center} 
\caption{Franson interferometry.}
\label{utransform}
\end{figure}

Under local unitary operations, the quantum state of the three-qubit photon 
pairs evolve keeping their entanglement unaltered. The 
topological nature of the phase evolution is strongly dependent on entanglement, 
so that it is important to identify entanglement signatures on the state 
evolution. As in Ref. \cite{topoluff}, signatures of entanglement can be found 
on interference patterns between the evolved and the initial state. 
Two-photon interference can be achieved with the well known Franson setup, 
where each photon from a quantum correlated pair is sent through 
two alternative paths, a long and a short one \cite{franson1,franson2,ryff,bhess}. 
When the delay 
time between the short and the long paths is larger than the detection time 
window, the two-photon coincidence count exhibits interference patterns. 
Each coincidence count may result from both photons following either 
the short or the long paths. Photons going through different 
paths do not coincide. Moreover, each arm of an SPDC source has a 
considerably short coherence length, so that no single photon 
interference can occur. Then, the overlap between the evolved and 
the initial state appears as the fringe visibility when the 
three qubits are individually operated in one arm and left unchanged 
in the other, as sketched in Fig. \ref{utransform}. 

\begin{figure}
\begin{center} 
\includegraphics[scale=.7]{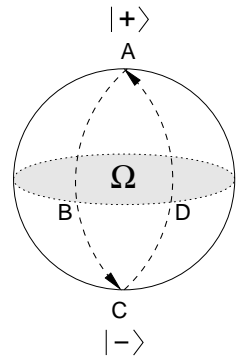}\\
\end{center} 
\caption{Paths on the Poincar\'e sphere.}
\label{poincare}
\end{figure}

In order to simplify the experimental proposal, still being able to cover the 
topological structure of the three-qubit local SU(2) orbits, we shall be dealing 
with diagonal unitary operations in the $\{\ket{+},\ket{-}\}$ basis. For each 
polarization transformation, a sequence of two HWPs (DHWP) oriented at $0$ and $\phi$, 
respectively, can be used. The same kind of diagonal transformation for OAM 
(LG basis) can be performed by a sequence of two DPs (DDP) at different orientations. 
These schemes are sketched in Fig. \ref{utransform}. 
It is instructive to view the operations performed by the DHWPs and DDP 
in the single qubit Poincar\'e representation. Starting from a $\ket{+}$ state, 
a HWP with its fast axis oriented at angle $\theta$ makes the transformation 
$\ket{+}\rightarrow\ket{\theta+\pi/4}\rightarrow\ket{-}$, where $\ket{\theta}$ 
represents a linear polarization state along a direction rotated by the angle 
$\theta$ with respect to the horizontal. Therefore, a sequence of two HWPs 
oriented at $\theta$ and $\theta + \phi$, respectively, makes the cycle 
$\ket{+}\rightarrow\ket{\theta+\pi/4}\rightarrow\ket{-}\rightarrow
\ket{\theta+\phi+\pi/4}\rightarrow\ket{+}$, which corresponds to 
path $A\rightarrow B\rightarrow C\rightarrow D\rightarrow A$ in 
Fig. \ref{poincare}. Since this path is composed of two geodesic segments, 
enclosing a solid angle $\Omega=\phi$, a purely geometric phase $\phi/2$ is acquired 
by single qubits initially prepared at $\ket{+}$. Of course, the same solid would be 
enclosed by an initial state $\ket{-}$, but in opposite direction, giving a geometric 
phase $-\phi/2$. Therefore, each degree of freedom follows the local $SU(2)$ operation:
\begin{equation}
U(\phi) = \left[
\begin{matrix}
e^{i\phi/2} & 0\\
0 & e^{-i\phi/2}
\end{matrix}
\right]\;,
\label{uphi}
\end{equation}
given in the $\{|+\rangle,|-\rangle\}$ basis, so that the overall three-qubit state will 
be transformed according to 
$U(\phi_s)\otimes U(\phi_o)\otimes U(\phi_i)\,$, where $\phi_s\,$, $\phi_o\,$, 
and $\phi_i$ correspond to signal polarization, OAM, and idler 
polarization, respectively. 

Under these local operations, product and entangled states evolve differently 
in what regards the overlap between the initial and the evolved states: 
$\bra{\psi}U(\phi_s)\otimes U(\phi_o)\otimes U(\phi_i)\ket{\psi}\,$. 
In order to access these differences and investigate the role played by 
entanglement and its relationship to the three-qubit topological structure, 
we must perform interferometric measurements where a proper background 
for comparison between product and entangled states can be established. 
A possible strategy is suggested in Fig. \ref{utransform}, where dynamical 
phases $\theta_s$ and $\theta_i$ are deliberately added in one arm of 
the interferometer. As these dynamical phases are continuously varied, 
the coincidence count exhibits an interference pattern that should evolve 
as the single qubit unitary operations are applied. In fact, the 
coincidence count is proportional to 
\begin{equation}
C=C_0\left[1 + \mathcal{V}
\cos(\theta + \Phi)\right]\;,
\label{pattern}
\end{equation}
where $C_0$ is the coincidence offset, $\ket{\psi}$ is one of the selected 
states discussed above, $\theta=\theta_s+\theta_i$ is the total dynamical phase 
added and 
\begin{eqnarray}
\Phi &\equiv &\arg\left[\bra{\psi}U(\phi_s)\otimes U(\phi_o)\otimes U(\phi_i)\ket{\psi}\right],\nonumber\\
\mathcal{V} &\equiv &|\bra{\psi}U(\phi_s)\otimes U(\phi_o)\otimes U(\phi_i)\ket{\psi}|\;.
\end{eqnarray}
Therefore, the absolute value of the overlap between the initial and the evolved states $\mathcal{V}$
is related to the fringe visibility, while the overlap phase $\Phi$, i.e., the Pancharatnam relative phase \cite{panch,recha}, translates to a fringe 
displacement. For a cyclic evolution, the fringes should recover maximal visibility 
and exhibit the accumulated phase shift, which is of topological nature 
for entangled states. However, the role of entanglement must be captured from signatures 
on the evolution of the interference pattern, as the individual unitary operations are 
applied. We shall investigate these signatures numerically in the next section. 

\section{Numerical Results}
\label{numerical}

To demonstrate the presence of the topological phases of the two different topological 
structures represented by the X state and the biased GHZ state we give examples of cyclic 
unitary evolutions in each homotopy class for both states. The evolution of the interference 
pattern for these entangled states, as the unitary evolution is gradually implemented, is 
compared to the evolution of the interference patterns of product states with the same local 
statistics in the $\{\ket{+},\ket{-}\}$ basis. Since we consider unitary evolutions that are 
diagonal in the $\{\ket{+},\ket{-}\}$ basis, no difference in the interference patterns 
between an entangled state and such a product state can be attributed to the local degrees 
of freedom. Any difference is thus due to entanglement.

\subsection{X state}

A set of cyclic evolutions of the X state resulting in a $\frac{\pi}{2}$ phase shift is generated by the unitary operators 
$U(\phi_{s}(t))\otimes{U(\phi_{o}(t))}\otimes{U(\phi_{i}(t))}$, where $t\in[0,T]$, 
such that $\phi_{s}(0)=\phi_{o}(0)=\phi_{i}(0)=0$ and $\phi_{s}(T)=\phi_{o}(T)=\phi_{i}(T)=-\pi$. 
For such an evolution of $|\psi_{X}\rangle$, the coincidence intensity $C$ as a function of  
$\theta,\phi_{s},\phi_{o}$, and $\phi_{i}$ is given by the expression

\begin{eqnarray}
C&=&C_{0}\bigg[1+\frac{1}{2}\cos\left(\theta+\frac{\phi_{s}}{2}\right)\cos\left(\frac{\phi_{o}+\phi_{i}}{2}\right)\nonumber\\
&&+\frac{1}{2}\cos\left(\theta-\frac{\phi_{s}}{2}\right)\cos\left(\frac{\phi_{o}-\phi_{i}}{2}\right)\bigg].\nonumber\\
\label{c0}
\end{eqnarray}

\begin{figure*}[ht]
\centering
\includegraphics[width=0.45\textwidth]{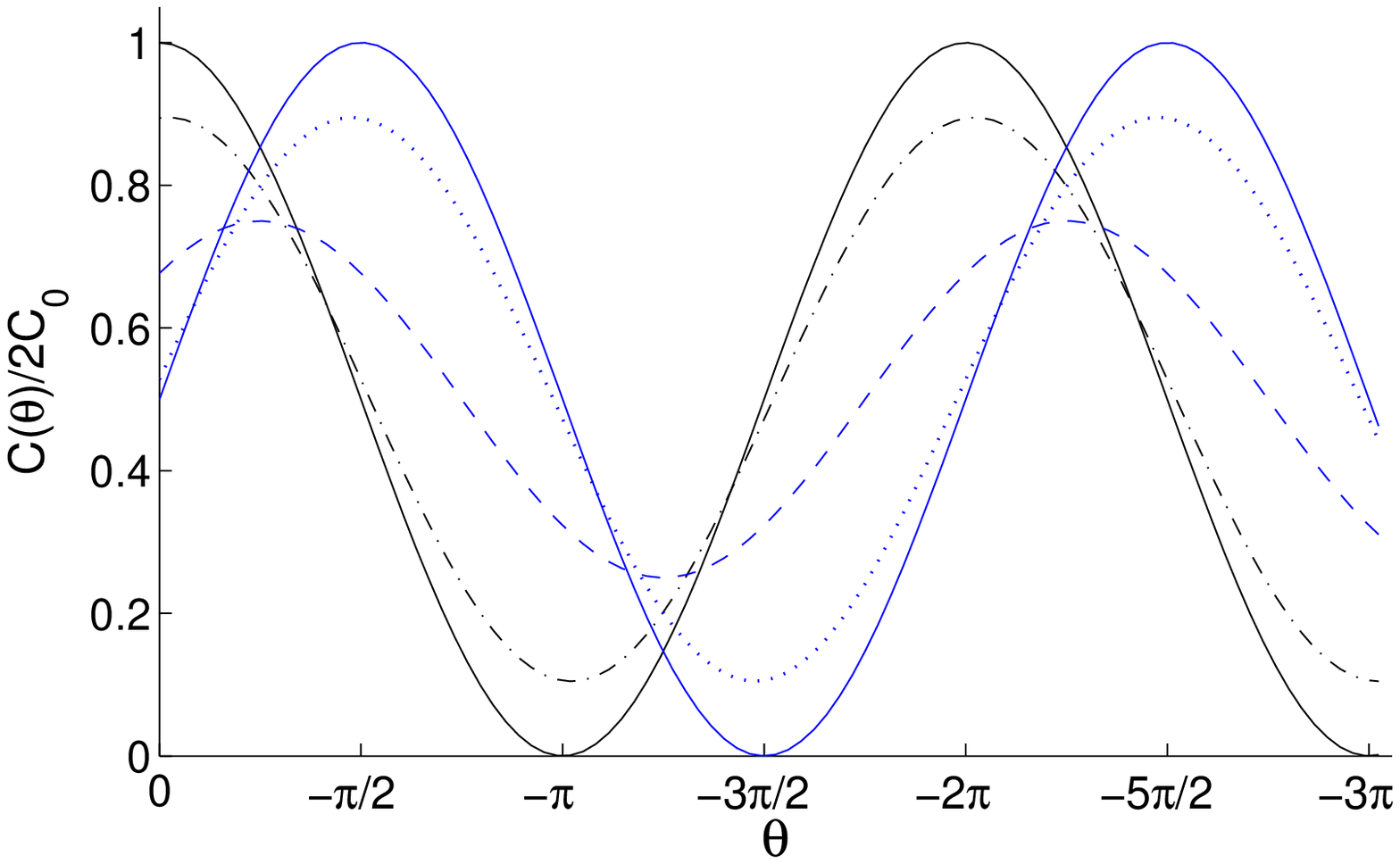}\phantom{hhhhh}
\includegraphics[width=0.45\textwidth]{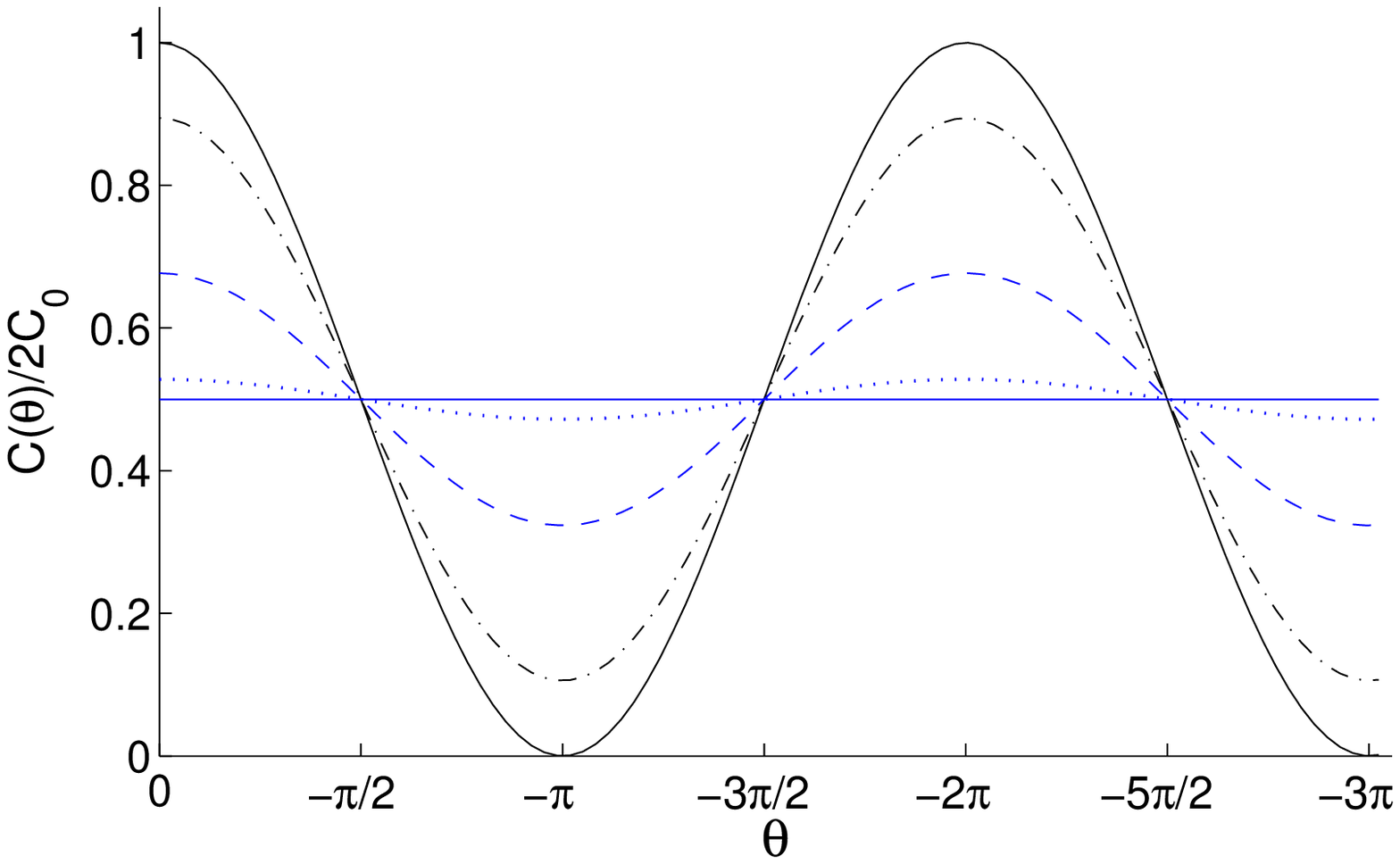}\phantom{.}
\caption{(Color online) The coincidence intensity $C(\theta)$ as a function of $\theta$ for 
selected values of $\frac{t}{T}$ of the evolution generated by $U_{X1}(t)$ for $|\psi_{X}\rangle$ 
(left) and $|\psi_{prod}\rangle$ (right). The different interference intensities correspond to 
the values $\frac{t}{T}=0$ (solid black), $\frac{t}{T}=\frac{1}{4}$ (dash-dotted black), 
$\frac{t}{T}=\frac{1}{2}$ (dashed blue), $\frac{t}{T}=\frac{3}{4}$ (dotted blue), and 
$\frac{t}{T}=1$ (solid blue), respectively. }
\label{fig:balgor4}
\end{figure*}

One unitary operator of this kind is $U_{X1}(t)$ given by $\phi_{s}(t)=\phi_{o}(t)=\phi_{i}(t)=-\pi{t}/T$.
If the X state is evolved by $U_{X1}(t)$ the coincidence intensity as a function of $t$ and $\theta$ is given by

\begin{eqnarray}
C(t,\theta)=C_{0}\left[1+\frac{1}{4}\cos\left(\theta-\frac{3\pi{t}}{2T}\right)+
\frac{3}{4}\cos\left(\theta+\frac{\pi{t}}{2T}\right)\right].\nonumber\\
\label{c1}
\end{eqnarray}
We can see that there is a reappearance of maximal fringe visibility for $\frac{t}{T}=1$ 
with the expected fringe shift $\frac{\pi}{2}$. Moreover, there are no values of $\frac{t}{T}$ 
for which the interference fringes disappear. This illustrates that, in contrast to the case 
of maximally entangled two-qubit states \cite{milman}, nontrivial topological phases can be 
obtained without going through a state orthogonal to the initial one. 
The coincidence intensity in Eq. (\ref{c1}) for selected values of $\frac{t}{T}$ is shown 
in the left panel of Fig. \ref{fig:balgor4}.

\begin{figure*}[ht]
\centering
\includegraphics[width=0.45\textwidth]{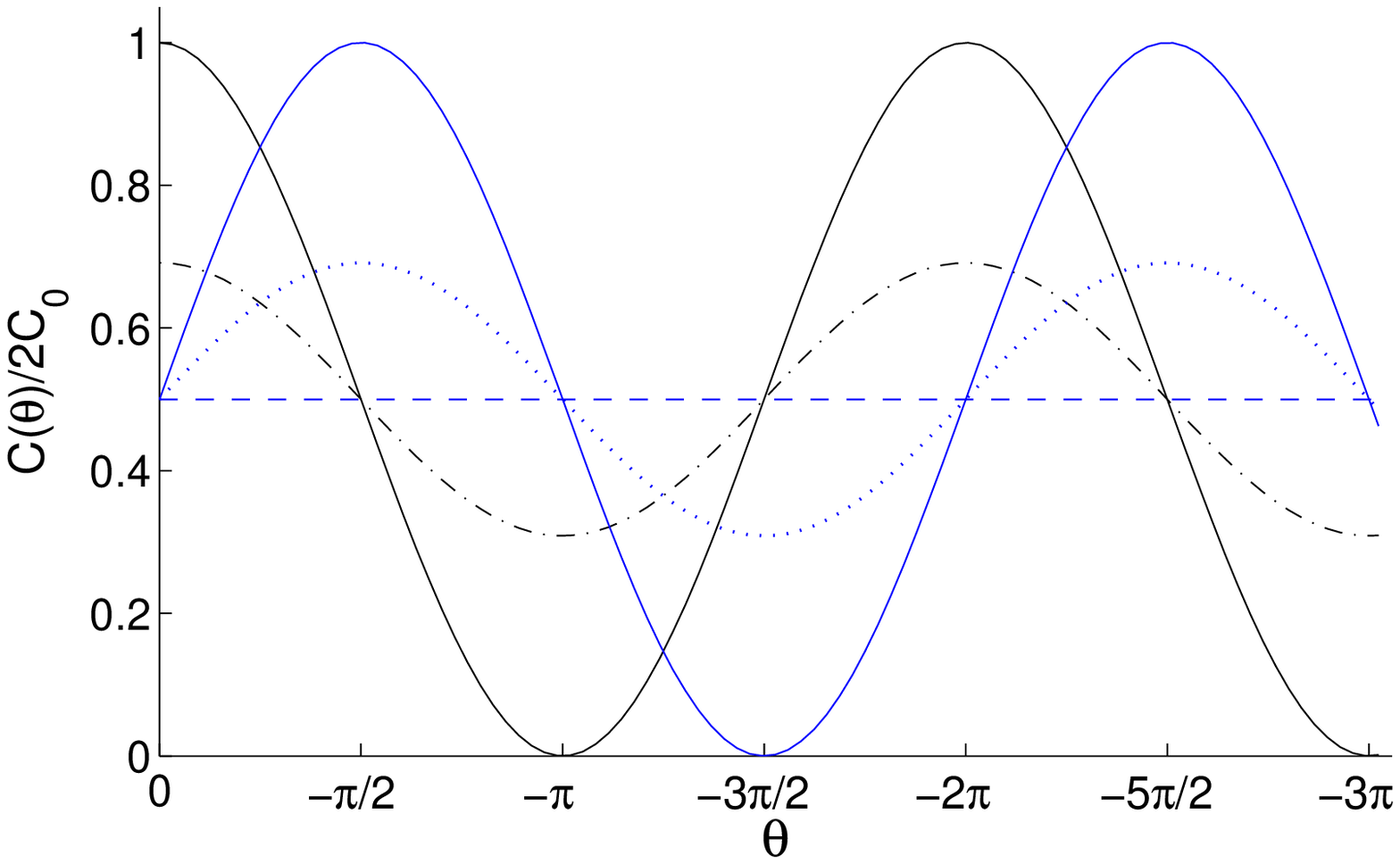}\phantom{hhhhh}
\includegraphics[width=0.45\textwidth]{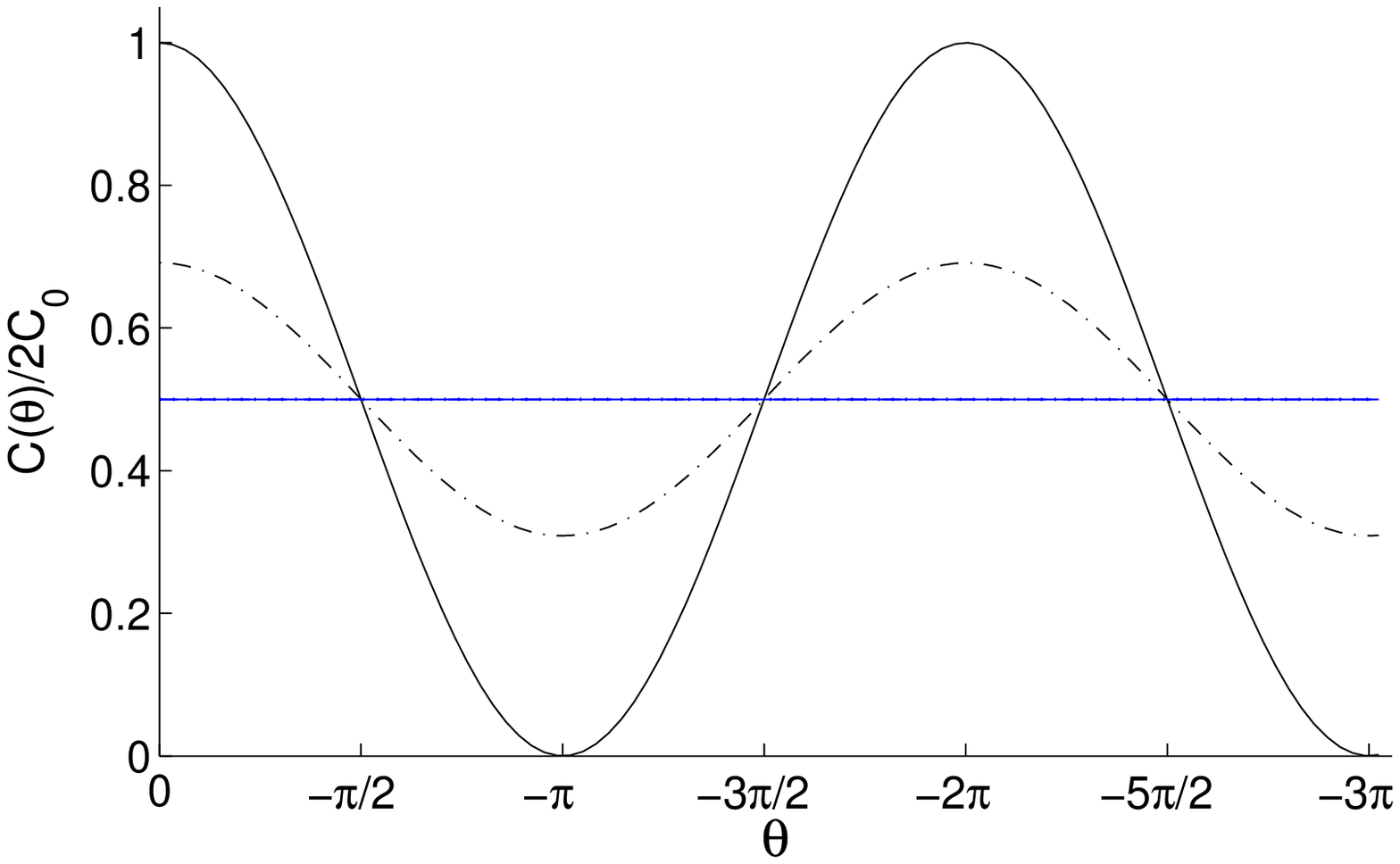}\phantom{.}
\caption{(Color online) The coincidence intensity $C(\theta)$ as a function of $\theta$ for 
selected values of $\frac{t}{T}$ of the evolution generated by $U_{X2}(t)$ for $|\psi_{X}\rangle$ 
(left) and $|\psi_{prod}\rangle$ (right). The different interference intensities correspond to 
the values $\frac{t}{T}=0$ (solid black), $\frac{t}{T}=\frac{1}{4}$ (dash-dotted black), 
$\frac{t}{T}=\frac{1}{2}$ (dashed blue), $\frac{t}{T}=\frac{3}{4}$ (dotted blue), and 
$\frac{t}{T}=1$ (solid blue), respectively. }
\label{fig:balgor5}
\end{figure*}

Another cyclic evolution in the same homotopy class can be generated by the unitary operator 
$U_{X2}(t)$ given by

\begin{eqnarray}
\phi_{s}(t)\!&=&\!-\pi-\pi\!\left[\frac{3{t}}{T}-1\right]H\!\left[\frac{1}{3}-\frac{t}{T}\right]\nonumber\\
\phi_{o}(t)\!&=&\!-\pi-\pi\!\left[\left(\frac{3{t}}{T}-1\right)H\!\left(\frac{t}{T}-\frac{1}{3}\right)-1\right]H\!\left[\frac{2}{3}-\frac{t}{T}\right]\nonumber\\
\phi_{i}(t)\!&=&\!-\pi-\pi\!\left[\left(\frac{3{t}}{T}-2\right)H\!\left(\frac{t}{T}-\frac{2}{3}\right)-1\right],\nonumber\\
\end{eqnarray}
for $0\leq{t}\leq{T}$, where $H$ is the Heaviside step function defined by $H(x)=0$ for 
$x<0$ and $H(x)=1$ for $x>0$. The coincidence intensity as a function of $\theta$ and 
$t$ in this case is

\begin{eqnarray}
C(t,\theta)&=&C_{0}\bigg[1+H\left(1-\frac{3t}{T}\right)\cos \theta \cos\left(\frac{3\pi{t}}{2T}\right)\nonumber\\
&&+H\left(\frac{3t}{T}-2\right)\sin \theta \sin\left(\frac{3\pi{t}}{2T}\right)\bigg].\nonumber\\
\label{c2}
\end{eqnarray}
Again we can see the expected fringe shift $\frac{\pi}{2}$ with maximal fringe visibility for 
$\frac{t}{T}=1$. For the cyclic evolution generated by $U_{X2}(t)$, as opposed to that generated 
by $U_{X1}(t)$,  the interference fringes disappear for $\frac{1}{3}\leq\frac{t}{T}\leq\frac{2}{3}$, 
meaning that the evolution takes the system through states orthogonal to the initial state during 
the evolution. With respect to the evolution of the fringe visibility there are thus qualitatively 
different evolutions in the same homotopy class. 
The coincidence intensity in Eq. (\ref{c2}) for selected values of $\frac{t}{T}$ is shown in 
the left panel of Fig. \ref{fig:balgor5}.

To verify that the fringe shifts are due to entanglement we consider the product state 
$\ket{\psi_{prod}}$, defined in Eq. (\ref{psiprod}), which has the same probability 
distributions for the local degrees of freedom as the X state in both the $\{\ket{H},\ket{V}\}$ 
and the $\{\ket{+},\ket{-}\}$ bases.  
The coincidence intensity as a function of $\theta,\phi_{s},\phi_{o}$, and $\phi_{i}$ for 
$\ket{\psi_{prod}}$, given that it is subjected to a unitary 
$U(\phi_{s})\otimes{U(\phi_{o})}\otimes{U(\phi_{i})}$, is

\begin{eqnarray}
C=C_{0}\left[1+\cos \theta \cos\left(\frac{\phi_{s}}{2}\right)
\cos\left(\frac{\phi_{o}}{2}\right)\cos\left(\frac{\phi_{i}}{2}\right)\right].
\nonumber\\
\label{cp}
\end{eqnarray}
We can see that fringe visibility for $\phi_{s}(T)=\phi_{o}(T)=\phi_{i}(T)=-\pi$ is zero. 
For $\ket{\psi_{prod}}$ the only values of $\phi_{s},\phi_{o}$, and $\phi_{i}$ that gives 
maximal fringe visibility are $0$ and $2\pi$.
Thus, the reappearance of maximal fringe visibility for the value $-\pi$ of 
$\phi_{s},\phi_{o}$ and $\phi_{i}$ with a 
$\frac{\pi}{2}$ fringe shift, is due to the entanglement of the X state. 
The coincidence intensity in Eq. (\ref{cp}) for the evolutions of $\ket{\psi_{prod}}$ 
generated by $U_{X1}(t)$ and by $U_{X2}(t)$ at selected values of $\frac{t}{T}$ is 
shown in the right panel of Figs. \ref{fig:balgor4} and \ref{fig:balgor5} respectively.

Note that the cyclic unitary evolutions that give maximal fringe visibility 
for $|\psi_{prod}\rangle$ are also cyclic evolutions of the X state. 
This however holds only for the diagonal unitary operators we are considering. 
A more general cyclic evolution of $|\psi_{prod}\rangle$ is not typically a cyclic 
evolution of $|\psi_{X}\rangle$.

The set of cyclic evolutions generating the $\pi$-phase shift of the X state that 
are diagonal in the $\{\ket{+},\ket{-}\}$ basis are the evolutions 
$U(\phi_{s}(t))\otimes{U(\phi_{o}(t))}\otimes{U(\phi_{i}(t))}$ $t\in[0,T]$ such 
that $\phi_{s}(0)=\phi_{o}(0)=\phi_{i}(0)=0$ and $\phi_{s}(T)=\phi_{o}(T)=\phi_{i}(T)=-2\pi$. 
However since these are also cyclic evolutions of the product state $|\psi_{prod}\rangle$ the 
reappearance of maximal fringe visibility and $\pi$ phase shift cannot be attributed to the 
entanglement of the X state. 

To observe a $\pi$ phase shift that cannot be attributed to local degrees of freedom we must 
implement a cyclic evolution generated by unitaries that are not diagonal in the 
$\{\ket{+},\ket{-}\}$ basis. We recall however that the X state is identical to the GHZ 
state in a different basis. 
For the GHZ state there are evolutions generated by diagonal unitaries leading to a $\pi$ phase 
shift that can be attributed to entanglement.

\subsection{GHZ and biased GHZ state}

\begin{figure*}[ht]
\centering
\includegraphics[width=0.45\textwidth]{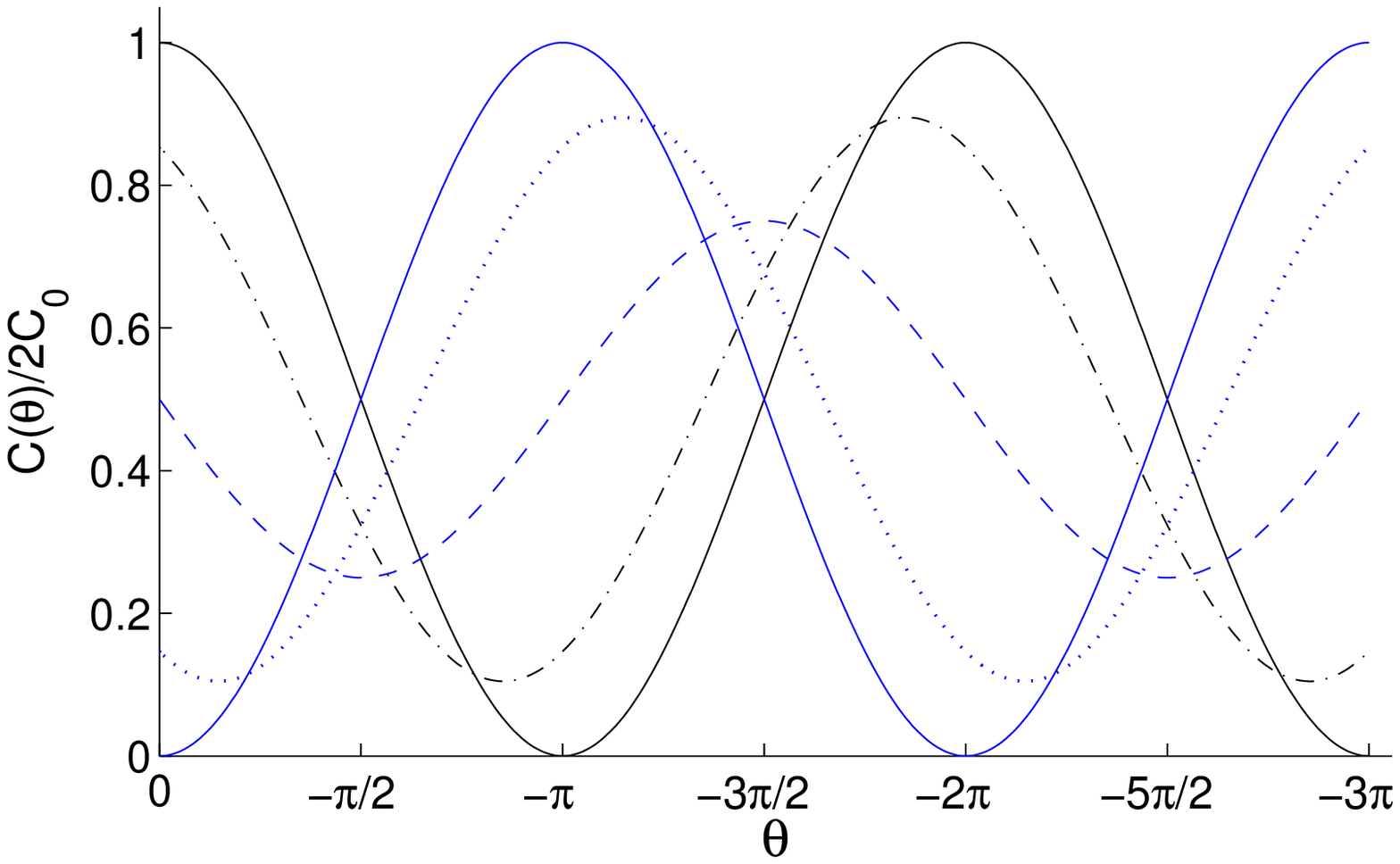}\phantom{hhhhh}
\includegraphics[width=0.45\textwidth]{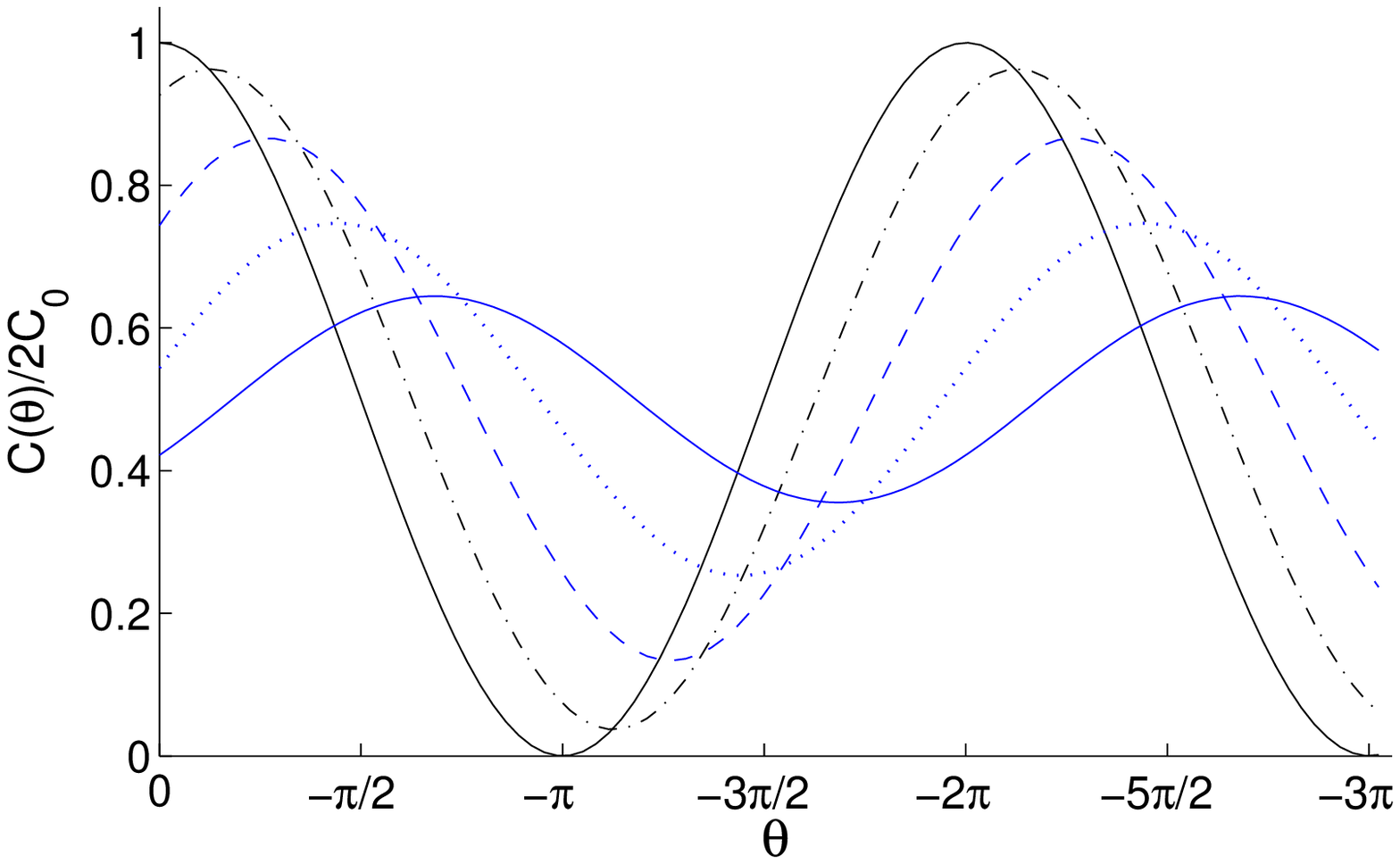}\phantom{.}
\centering
\includegraphics[width=0.45\textwidth]{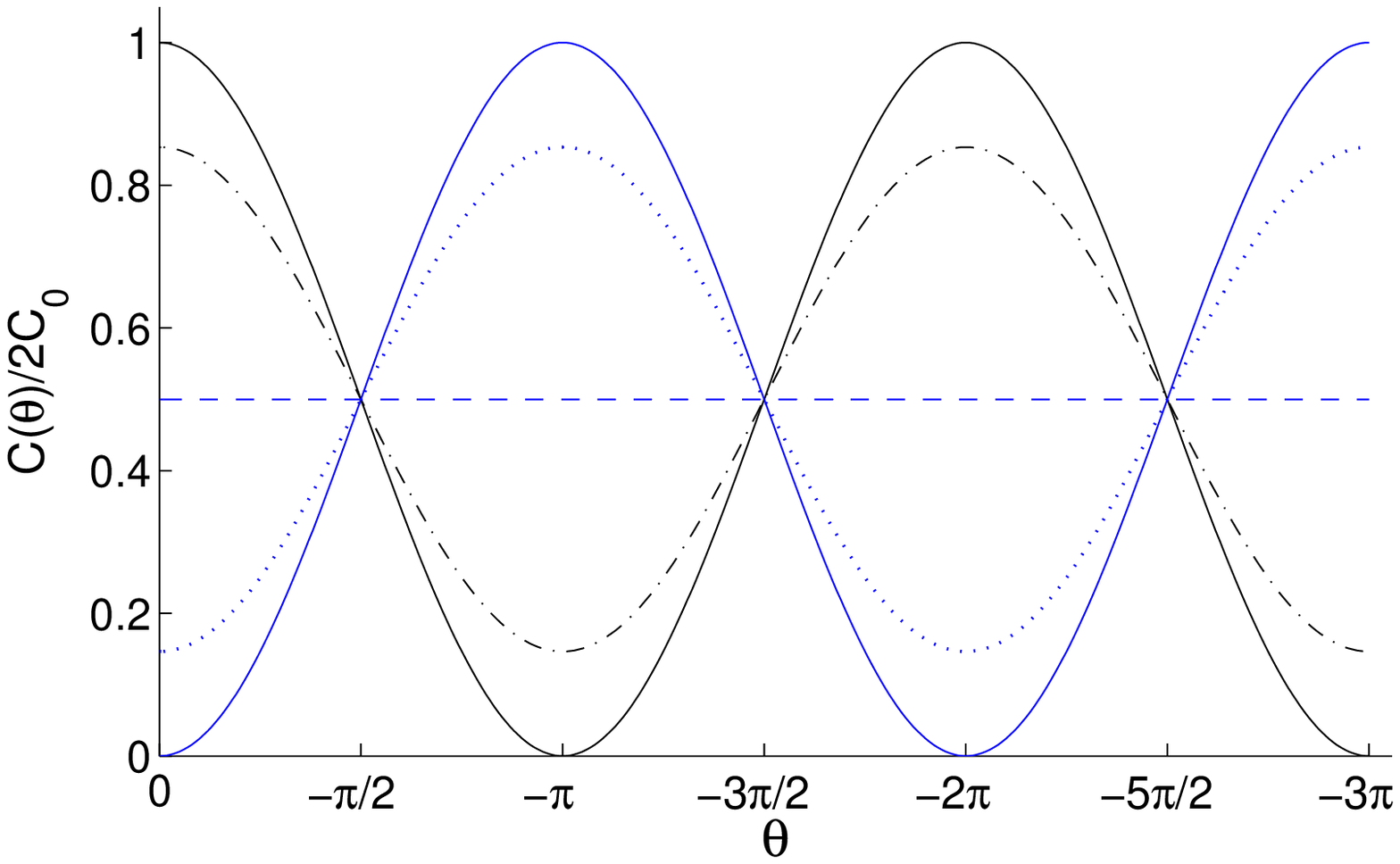}\phantom{hhhhh}
\includegraphics[width=0.45\textwidth]{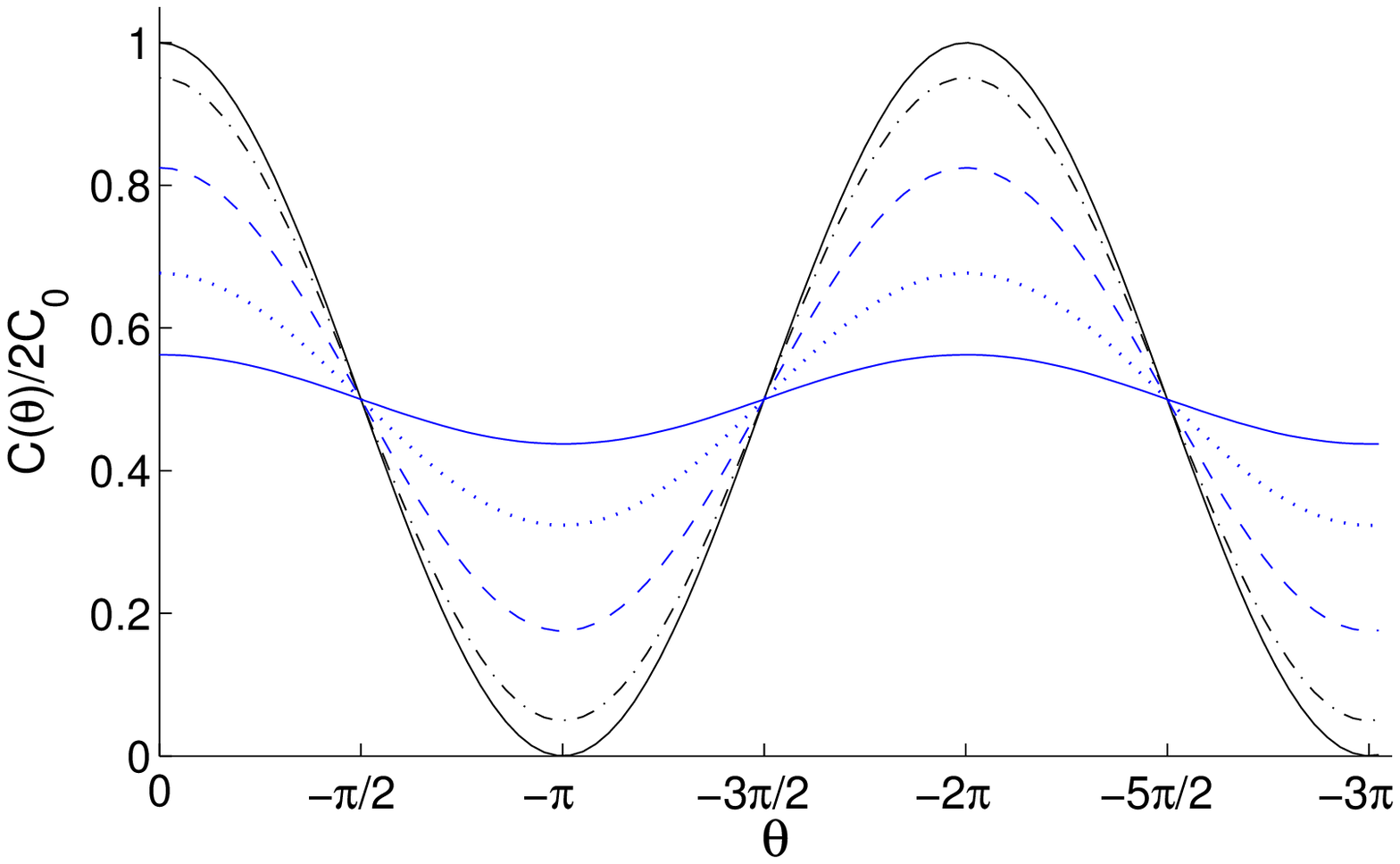}\phantom{.}
\caption{(Color online) The coincidence intensity $C(\theta)$ as a function of 
$\theta$ for selected values of $\frac{t}{T}$ of the evolution generated by $U_{bghz}(t)$ 
for $|\psi_{bghz}\rangle$ (upper left), $|\psi^{\prime}_{prod}\rangle$ (upper right), 
$|\psi_{ghz}\rangle$ (lower left), and $|\psi_{prod}\rangle$ (lower right). The different 
interference intensities correspond to the values $\frac{t}{T}=0$ (solid black), 
$\frac{t}{T}=\frac{1}{4}$ (dash-dotted black), $\frac{t}{T}=\frac{1}{2}$ (dashed blue), 
$\frac{t}{T}=\frac{3}{4}$ (dotted blue), and $\frac{t}{T}=1$ (solid blue), respectively.}
\label{fig:balgor3}
\end{figure*}

We consider the GHZ state $\ket{\psi_{ghz}}=\frac{1}{\sqrt{2}}|+++\rangle+\frac{1}{\sqrt{2}}|---\rangle$ and 
the biased GHZ state $\ket{\psi_{bghz}}=\frac{1}{2}|+++\rangle+\frac{\sqrt{3}}{2}|---\rangle$.
The GHZ state and the biased GHZ state share a set of cyclic evolutions that give rise to 
a $\pi$ phase shift and are diagonal in the $\{\ket{+},\ket{-}\}$ basis. We recall however 
that the GHZ state and the biased GHZ state are representatives of different classes of 
topological phase structures of the local SU(2) orbits since for the GHZ state there are cyclic 
evolutions resulting in a $\frac{\pi}{2}$ phase shift while for the biased GHZ state there 
are no such evolutions.

The cyclic evolutions of these states that result in a $\pi$ phase shift are generated by 
unitaries $U(\phi_{s}(t))\otimes{U(\phi_{o}(t))}\otimes{U(\phi_{i}(t))}$, where $t\in[0,T]$, 
such that $\phi_{s}(0)=\phi_{o}(0)=\phi_{i}(0)=0$, and $\phantom{yy}\phi_1(T)+\phi_2(T)+\phi_3(T)=\pm{2}\pi$.
There are no evolutions generated by unitaries of this kind that takes the biased GHZ state 
through a state orthogonal to the initial state. Thus, the interference fringes never disappears. 
The GHZ state on the other hand is evolved through an orthogonal state.

One unitary operator of this kind is $U_{bghz}(t)$ given by the choice 
$\phi_{s}(t)=\phi_{o}(t)=\phi_{i}(t)=\frac{2\pi{t}}{3T}$.
The coincidence intensity for the evolution of $|\psi_{ghz}\rangle$ generated by $U_{bghz}(t)$ 
will be a function of $t$ and $\theta$ given by

\begin{eqnarray}
C(t,\theta)=C_{0}\left[1+\cos \theta \cos\left(\frac{\pi{t}}{T}\right)\right].
\label{c31}
\end{eqnarray}
The coincidence intensity in Eq. (\ref{c31}) for selected values of $\frac{t}{T}$ is shown 
in the lower left panel of Fig. \ref{fig:balgor3}.
The coincidence intensity for the evolution of $|\psi_{bghz}\rangle$ generated by 
$U_{bghz}(t)$ on the other hand will include an additional term and is given as a 
function of $t$ and $\theta$ by

\begin{eqnarray}
C(t,\theta)=C_{0}\left[1+\cos \theta \cos\left(\frac{\pi{t}}{T}\right)-\frac{1}{2}\sin \theta 
\sin\left(\frac{\pi{t}}{T}\right)\right].\nonumber\\
\label{c3}
\end{eqnarray}
The coincidence intensity in Eq. (\ref{c3}) for selected values of $\frac{t}{T}$ is 
shown in the upper left panel of Fig. \ref{fig:balgor3}.
For both the GHZ state and the biased GHZ state the coincidence intensity depends 
only on the sum $\phi_{s}(t)+\phi_{o}(t)+\phi_{i}(t)$ and not on the $\phi_{i}(t)$ 
individually.  For both states there is a reappearance of maximal fringe visibility 
at $\frac{t}{T}=1$ and the interference fringes are shifted by a $\pi$ phase.

To see the signature of the entanglement in the interference pattern of 
$|\psi_{bghz}\rangle$, we also consider the product state

\begin{eqnarray}
\ket{\psi^{\prime}_{prod}}&=&
\left(\frac{1}{2}\ket{+}+\frac{\sqrt{3}}{2}\ket{-}\right)\otimes
\left(\frac{1}{2}\ket{+}+\frac{\sqrt{3}}{2}\ket{-}\right)
\nonumber\\
&\otimes&
\left(\frac{1}{2}\ket{+}+\frac{\sqrt{3}}{2}\ket{-}\right)\;,
\label{psiprodprimes}
\end{eqnarray}
which has the same probability distributions for the local degrees of freedom in the 
$\{\ket{+},\ket{-}\}$ basis as $|\psi_{bghz}\rangle$.
The interference intensity as a function of $\theta$ and $t$ when 
$|\psi^{\prime}_{prod}\rangle$ is evolved by $U_{bghz}(t)$ is 

\begin{eqnarray}
C(t,\theta)&=&C_{0}\bigg\{1+\cos \theta \cos\left[\frac{\pi{t}}{3T}\right]
\left[1-\frac{7}{4}\sin^2\left(\frac{\pi{t}}{3T}\right)\right]\nonumber\\
&&+\frac{3}{2}\sin \theta \sin\left[\frac{\pi{t}}{3T}\right]
\left[1-\frac{13}{12}\sin^2\left(\frac{\pi{t}}{3T}\right)\right]\bigg\}.
\nonumber\\
\label{c3p}
\end{eqnarray}

Comparing Eqs. (\ref{c3}) and (\ref{c3p}) we see that the reappearance of 
maximal fringe visibility for $\frac{t}{T}=1$ is absent for $|\psi^{\prime}_{prod}\rangle$. 
Moreover the fringe shift is not equal to $\pi$. Thus, the reappearance of fringe 
visibility and $\pi$ phase shift is thus due to entanglement. 
The coincidence intensity for $|\psi^{\prime}_{prod}\rangle$, evolved by $U_{bghz}$, is 
shown for selected values of $\frac{t}{T}$ in upper right panel of Fig. \ref{fig:balgor3} 
alongside the coincidence intensity of $|\psi_{bghz}\rangle$.

To see the signature of entanglement in the interference pattern of $|\psi_{ghz}\rangle$ 
in Eq. (\ref{c31}) we compare with the interference patterns of $|\psi_{prod}\rangle$. 
The coincidence intensity for $|\psi_{prod}\rangle$ is given by Eq. (\ref{cp}) and we 
can see that there is no reappearance of maximal fringe visibility for 
$\phi_{s}(t)=\phi_{o}(t)=\phi_{i}(t)=\frac{2\pi}{3}$. Thus, the reappearance for 
$|\psi_{bghz}\rangle$ can be attributed to entanglement and the phase shift is of topological nature. The coincidence intensity 
for $|\psi_{prod}\rangle$, evolved by $U_{bghz}$, is shown for selected values of 
$\frac{t}{T}$ in the lower right panel of Fig. \ref{fig:balgor3} alongside the 
coincidence intensity for $|\psi_{ghz}\rangle$.

\section{Conclusions}

We propose an experimental scheme to observe the topological phases acquired by 
special classes of three-qubit states. These phases reveal the nontrivial 
topological structure of the local SU(2) orbits. In particular, observation of the $\pi/2$ topological phase shift would be a signature
of multiqubit entanglement as this phase exists only for more than two 
qubits. The experimental proposal is within 
the technological resources available in most quantum optics laboratories, 
and can be implemented in a short term. Furthermore, the insensitivity to continuous path deviations of the unitary evolution is a robust feature of the topological phases with potential applications to 
quantum information processing.

\acknowledgments
M.J., E.S., and K.S. acknowledge support from the National Research Foundation and the Ministry of
Education (Singapore).
A. Z. Khoury acknowledges funding from the Brazilian agencies 
Coordena\c c\~{a}o de 
Aperfei\c coamento de Pessoal de N\'\i vel Superior (CAPES), 
Funda\c c\~{a}o de Amparo \`{a} Pesquisa do Estado do Rio de Janeiro 
(FAPERJ-BR), and Conselho Nacional de Desenvolvimento Cient\'{\i}fico e 
Tecnol\'ogico (CNPq). A. Z. Khoury is deeply grateful for the hospitality 
during the visit to Centre for Quantum Technologies at the National
 University of Singapore, Singapore.

\end{document}